\documentclass[11pt,twoside,smallextended]{article}
\usepackage{amsfonts,epsfig}
\usepackage{amsthm}
\usepackage{xy}
\input{xy}
\xyoption{all} \xyoption{rotate}

\theoremstyle{remark}

\newtheorem{remark}{Remark}[subsection]

\newtheorem{definition}{Definition}[subsection]

\theoremstyle{plain}

\newtheorem{proposition}{Proposition}[subsection]

\begin{document}

\title{Homomorphisms of connectome graphs}
\author{
Peteris\ Daugulis\\ Department of Mathematics \\Daugavpils
University\\
Daugavpils, Latvia \\
peteris.daugulis@du.lv }

\maketitle

\begin{abstract} We propose to study homomorphisms of connectome
graphs.  Homomorphisms can be studied as sequences of elementary
homomorphisms - folds, which identify pairs of vertices. Several
fold types are defined. Initial computation results for some
connectome graphs are described.

\end{abstract}

\setcounter{tocdepth}{3} \tableofcontents

\section{Introduction} \label{intro-sec}

\subsection{Setup}

\subsubsection{The subject of study}

Connectome graphs are important discrete mathematical models used
for modelling nervous systems on different scales, see \cite{K},
\cite{S1}, \cite{S2}. In this paper we are mainly interested in
graph-theoretic problems and computations therefore we do not deal
with connectome scale and other modelling issues. For simplicity
of language we assume that we work on microscale - graph model
vertices are neurons (nerve cells) and edges are neuron
connections (e.g. synapses). Edges may be directed, undirected and
labelled. We use data available at
{http://www.openconnectomeproject.org}. We only consider the
directed graph structure of connectomes, the edge weights and
labels are not used.

%We list some problems currently being studied and described in
%literature:
%\begin{enumerate}
%
%\item detection of clusters, i.e. large subgraphs with extremal
%parameter (degree, clustering coeficient etc.) values,
%
%\item detection of vertices with special (extremal) properties
%(hubs, "rich clubs" etc.)
%
%\item detection of motifs, i.e. isomorphism types of subgraphs
%which occur relatively more often in connectomes than in random
%graphs.
%
%\end{enumerate}

\subsubsection{Graph homomorphisms}

We remind the reader basic definitions from graph theory, see
\cite{D}. A graph is a visualization of a relation - a pair
$\Gamma=(V,E)$, where $V$ is a set and $E\subseteq V\times V$. An
edge from vertex $u$ to vertex $v$ is a pair $[u,v]\in E$.
Directed graphs have ordered pairs. $\Gamma_{+}(v)$ is the set of
vertices having edges out of $v$, $\Gamma_{-}(v)$ is the set of
vertices having edges into $v$. If  $\Gamma=(V,E)$ is a graph and
there is a partition $\pi$ of $V$: $V=\bigcup_{i}V_{i}$, then a
\sl quotient graph mod $\pi$\rm\ denoted as $\Gamma/\pi$ is the
graph with vertex set $\{V_{i}\}$, two elements of the partition
are connected if and only if there is an edge between some two
vertices in the corresponding partition subsets.

We consider connectome graphs as being directed loopless graphs.

Given two graphs $\Gamma_{1}=(V_{1},E_{1})$ and
$\Gamma_{2}=(V_{2},E_{2})$, a function $f:\Gamma_{1}\rightarrow
\Gamma_{2}$ or $f:V_{1}\rightarrow V_{2}$ is called a \sl graph
homomorphism\rm\ provided $[u,v]\in E$ implies $[f(u),f(v)]\in E$.
In other words, edges are mapped to edges and their directions are
preserved. Any homomorphism is equivalent in a certain sense to a
homomorphism $\Gamma\rightarrow \Gamma/\pi$ for some partition
$\pi$ and thus can be thought as an identification of some
vertices and edges of $\Gamma$. The homomorphisms capture the
notion of maps preserving graph structure. Homomorphisms
$\Gamma\rightarrow \Gamma$ are called $\Gamma$-endomorphisms.
Noninjective homomorphisms preserve graph structure and decrease
the number of vertices. Thus homomorphisms may be used to analyze
big graphs such as connectome graphs.

An $\Gamma$-endomorphism $f:\Gamma\rightarrow \Gamma$ is called a
\sl retracting endomorphism\rm\ and $Im(\Gamma)$ - a retract,
provided $f|_{Im(\Gamma)}=id$. In other words, $Im(\Gamma)$ is
fixed by $f$ and $f\circ \iota=id$, where $\iota$ is the inclusion
of $Im(\Gamma)$ into $\Gamma$.

Another type of maps which preserve graphs structure are graph
isomorphisms. Isomorphisms are bijective maps - essentially
relabelings of vertices. It seems that relabelings can not
significantly help to understand structure of big graphs. In graph
theory relabelings are used to design normal forms of graphs.
Additionally, it is quite unlikely that two different connectome
graphs would have identical structure. It would interesting to
find such a biological phenomenon, but it is not within the scope
of our work. On the other hand, it is likely that homomorphic
images of certain connectome subgraphs are isomorphic (essential
structure of certain tissues etc. may be identical), thus both
homomorphisms and isomorphisms should be studied.

We call a graph homomorphism $f:\Gamma_{1}\rightarrow \Gamma_{2}$
\sl terminal\rm\ if there is no noninjective homomorphism
$\Gamma_{2}\rightarrow \Gamma_{3}$. In other words, terminal
homomorphisms cannot be composed with other noninjective
homomorphisms. Vaguely speaking, if we consider homomorphisms as
graph simplifications, the graph $\Gamma_{2}$ cannot be simplified
further. Images of terminal homomorphisms are called \sl terminal
graphs.\rm\

In case of undirected graphs terminal graphs are called \sl
cores.\rm\ The core of an undirected or directed graph is
determined uniquely up to isomorphism, see \cite{G}.

For undirected graphs terminal homomorphisms map graphs to
complete graphs. The minimal $c$ such that there is a homomorphism
$\Gamma\rightarrow K_{c}$ is called the chromatic number of
$\Gamma$.

 It is known that homomorphisms of undirected graphs
can be factored as sequences of \sl elementary homomorphisms\rm\
(which identify nonadjacent vertices) or \sl\ simple folds\rm\
(which identify nonadjacent vertices having a common neighbour),
see \cite{G}, \cite{H}. In this paper we define several types of
elementary homomorphisms of directed graphs.

\subsubsection{Main objectives and steps of our work}\label{12}

In the published literature connectome graphs are studied as fixed
objects. Some of the main directions of graph-theoretic study are
related to 1) probabilistic graph theory - connectome graphs are
compared to random graphs, 2) network analysis - various "network"
invariants, for instance, centrality invariants, are studied,
vertices having extremal invariant values ("rich club") are
identified, 3) small subgraph analysis - motifs characteristic for
connectomes are identified and 4) graph decompositions - subgraphs
having special properties are indentified.

We propose to study simplified connectome graphs - to study images
of connectome graphs under noninjective homomorphisms - maps
preserving adjacency of vertices and decreasing the number of
vertices. We also express noninjective homomorphisms as sequences
of vertex identifications. In other terms, we propose to study
quotient graphs of connectome graphs.

The main objective of our work is to compute terminal connectome
graph homomorphisms and endomorphisms for some connectome graphs
and discuss the results.

The main steps of our work: 1) definition of $5$ types of
elementary homomorphisms - folds, 2) computation of graph
invariants of the initial connectome, 3) computation of terminal
sequences of folds of each of the $5$ types, 4) computation of
graph invariants of the obtained terminal graphs.

\section{Main results}

\subsection{Types of elementary homomorphisms}

In this paper we introduce several types of elementary graph
homomorphisms - \sl folds\rm\ and study sequences of folds. A fold
is a graph homomorphism which identifies two vertices.
Furthermore, the following edges are identified: edges which have
common sources and point to identified vertices and edges which go
from identified vertices and have common targets. We propose the
following fold types which can be considered as connectome
simplification steps.
\begin{definition} A graph homomorphism is called a \sl forward
fold\rm\ if it identifies two nonadjacent vertices $u$ and $v$ of
$\Gamma$ (substitutes $u$ and $w$ by a new vertex $w$) such that
there is a vertex $x$ having properties $u\rightarrow x$ and
$v\rightarrow x$. It corresponds to the iden\-ti\-fi\-ca\-tion of
two neurons $u$ and $v$ which have a common synapse target neuron
$x$. See Fig.1.

$$
\xymatrix{
&x&&&&&x\\
&&&\ar[r]&&&\\
u\ar[uur]\ar@{.}[rr]&&v\ar[uul]&&&&w\ar[uu]\\
}
$$
\begin{center}

\

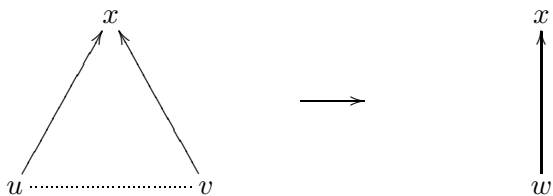
Fig.1.  - forward fold

\end{center}

\end{definition}

\begin{definition} A graph homomorphism is called a \sl
backward fold\rm\ if it identifies two nonadjacent vertices $u$
and $v$ of $\Gamma$ such that there is a vertex $y$ having
properties $y\rightarrow u$ and $y\rightarrow v$. It corresponds
to the identification of two neurons $u$ and $v$ which have a
common synapse source neuron $y$. See Fig.2.

$$
\xymatrix{
u\ar@{.}[rr]&&v&&&&w\\
&&&\ar[r]&&&\\
&y\ar[uul]\ar[uur]&&&&&y\ar[uu]\\
}
$$

\begin{center}

\

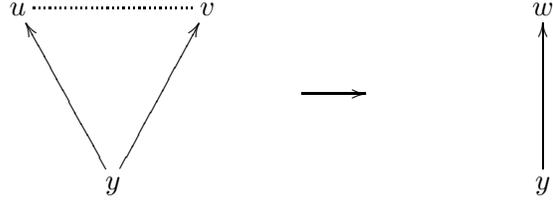
Fig.2.  - backward fold

\end{center}

\end{definition}

\begin{definition} A graph homomorphism is called a \sl
disjunctive fold\rm\ if it identifies two nonadjacent vertices $u$
and $v$ of $\Gamma$ such that there is either a vertex $x$ having
properties $u\rightarrow x$ and $v\rightarrow x$ or a vertex $y$
having properties $y\rightarrow u$ and $y\rightarrow v$. It
corresponds to the identification of two neurons $u$ and $v$ which
have either a common synapse target or source.
\end{definition}

\begin{definition} A graph homomorphism is called a \sl
conjunctive fold\rm\ if it identifies two nonadjacent vertices $u$
and $v$ of $\Gamma$ such that there is a vertex $x$ having
properties $u\rightarrow x$, $v\rightarrow x$, $x\rightarrow u$
and $x\rightarrow v$. It corresponds to the identification of two
neurons $u$ and $v$ which have a common synapse target and source
neuron $x$.
\end{definition}

\begin{definition} A graph homomorphism is called a \sl
retractive fold\rm\ if it identifies two nonadjacent vertices $u$
and $v$ of $\Gamma$ such that $\Gamma_{+}(u)\subseteq
\Gamma_{+}(v)$ and $\Gamma_{-}(u)\subseteq \Gamma_{-}(v)$. The
image of a retractive fold $\Gamma\rightarrow \Gamma'$ can be
thought as a subgraph of $\Gamma$, a composition of retractive
folds can be thought as a retractive endomorphism. See Fig.3. for
an example.

$$
\xymatrix{
x_{1}&x_{6}&&&&&x_{1}&x_{6}\\
x_{2}&&&&&&x_{2}&\\
x_{3}&u\ar[l]\ar[ul]\ar[uul]\ar[uu]\ar@{.}[rr]&&v\ar[uull]&\ar[r]&&x_{3}&w\ar[l]\ar[ul]\ar[uul]\ar[uu]\\
x_{4}\ar[ur]&&&&&&x_{4}\ar[ur]&\\
x_{5}\ar[uur]&x_{7}\ar[uu]\ar[uurr]&&&&&x_{5}\ar[uur]&x_{7}\ar[uu]\\
}
$$
\begin{center}

\

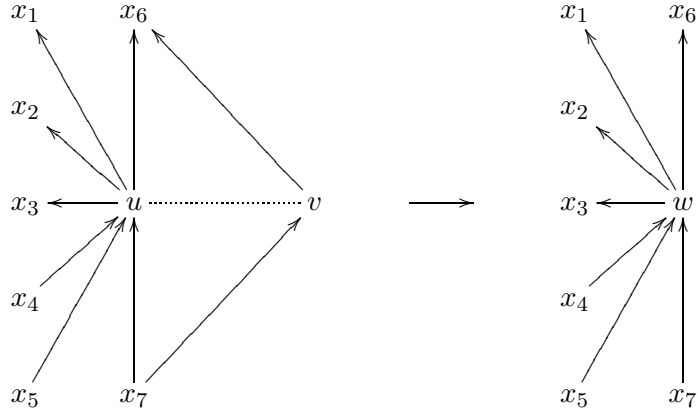
Fig.3.  - retractive fold

\end{center}

\end{definition}

\begin{remark} Computation of a sequence of retractive folds in an
arbitrary way does not guarantee that the terminal graph will be a
core.

\end{remark}

\begin{proposition}Images of folding
homomorphisms defined above have following properties:

\begin{enumerate}

\item  the only induced cycles of terminal graphs are oriented
cycles and (arbitrarily oriented) triangles;

\item folds preserve strong connectivity.

\end{enumerate}
\end{proposition}

We also propose a fold type which, apart from identifying
vertices, change edge weights. If two vertices are identified as
well as edges having a common vertex then it is natural to
increase weights of edges after identification. In this paper
there are no computational results for this fold type. To consider
these folds we assume that connectome graphs are weighted graphs -
edges have integer weights. Edge weight can be interpreted as
connection significance.

\begin{definition} Let $\Gamma$ be a weighted graph. A graph
homomorphism is called an \sl edge fold\rm\ if

\begin{enumerate}

\item it has type of one of the previously defined folds,

\item if the fold identifies vertices $u$ and $v$ creating a new
vertex $w$, then edges $u\stackrel{f}{\rightarrow} x$ and
$v\stackrel{g}{\rightarrow} x$ after identification give rise to
the edge $w\stackrel{f+g}{\rightarrow} x$, similarly for ingoing
edges.

\end{enumerate}

\end{definition}

\subsection{Fold sequences} Given a connectome graph and a chosen
fold type we perform a maximal sequence of consecutive folds until
no further fold is possible. The order of folds is arbitrary. In
the end we get a \sl terminal graph.\rm\

\subsection{Some examples}

In this subsection we describe some of our computational results.
All connectome graphs with number of vertices not exceeding $2000$
can be processed on a standard laptop computer.

\subsubsection{Cat}

Filename - Mixed.species\_brain\_1.graphml.

Initial graph description - strongly connected graph with $65$
vertices and $1139$ edges, underlying undirected graph has vertex
connectivity $6$, diameter $3$, radius $2$, center has $23$
vertices, minimal degree $3$, maximal degree $45$, degree sequence
$[ 0, 0, 0, 1, 1, 1,$ $ 0, 0, 0, 1, 3, 0, 4,$ $1, 2, 3, 2, 5,$ $1,
2, 2, 3, 2, 2, 3, 4$, $4$, $2$, $0$, $1$, $1$, $2,$  $1, 1, 1$,
$2, 1, 0, 1, 1, 0,$ $0, 1, 0, 1, 2 ]$.

Forward, backward, disjunctive and conjunctive folding terminal
graph description - strongly connected graph with $17$ vertices,
underlying undirected graph is the complete graph $K_{17}$.

Retract folding terminal graph description - strongly connected
graph with $61$ vertices, $1102$ edges, connectivity $9$, diameter
$3$, radius $2$, minimal degree $9$, maximal degree $44$.

%\begin{tabular}{|c|c|c|c|c|}
%  \hline
%  % after \\: \hline or \cline{col1-col2} \cline{col3-col4} ...
%  \#vertices & \#edges & connect.& \#weak comp. & \#big str.conn.comp.   \\
%  \hline
%  65 & 1139 & 3 & 1 & 1 \\
%  \hline
%\end{tabular}
%
%
%
%Forward fold
%
%\begin{tabular}{|c|c|c|c|c|}
%  \hline
%  % after \\: \hline or \cline{col1-col2} \cline{col3-col4} ...
%  \#vertices & \#edges & connect.& \#weak comp. & \#big str.conn.comp.   \\
%  \hline
%  17 & 269 & 16 & 1 & 1 \\
%  \hline
%\end{tabular}
%
%Backward fold
%
%\begin{tabular}{|c|c|c|c|c|}
%  \hline
%  % after \\: \hline or \cline{col1-col2} \cline{col3-col4} ...
%  \#vertices & \#edges & connect.& \#weak comp. & \#big str.conn.comp.   \\
%  \hline
%  17 & 269 & 16 & 1 & 1 \\
%  \hline
%\end{tabular}
%
%Disjunctive fold
%
%\begin{tabular}{|c|c|c|c|c|}
%  \hline
%  % after \\: \hline or \cline{col1-col2} \cline{col3-col4} ...
%  \#vertices & \#edges & connect.& \#weak comp. & \#big str.conn.comp.   \\
%  \hline
%  17 & 269 & 16 & 1 & 1 \\
%  \hline
%\end{tabular}
%
%Conjunctive fold
%
%\begin{tabular}{|c|c|c|c|c|}
%  \hline
%  % after \\: \hline or \cline{col1-col2} \cline{col3-col4} ...
%  \#vertices & \#edges & connect.& \#weak comp. & \#big str.conn.comp.   \\
%  \hline
%  65 & 1139 & 3 & 1 & 1 \\
%  \hline
%\end{tabular}
%
%Retract fold
%
%\begin{tabular}{|c|c|c|c|c|}
%  \hline
%  % after \\: \hline or \cline{col1-col2} \cline{col3-col4} ...
%  \#vertices & \#edges & connect.& \#weak comp. & \#big str.conn.comp.   \\
%  \hline
%  64 & 1133 & 4 & 1 & 1 \\
%  \hline
%\end{tabular}

\subsubsection{Worm}

Filename - c.elegans.herm\_pharynx\_1.graphml.

Initial graph description - $279$ vertices and $2993$ edges, $6$
strongly connected components - one with  $274$ vertices and the
other trivial components ($1$-vertex components), underlying
undirected graph has vertex connectivity $2$ ($5$ saparation
pairs, $11$ maximal $3$-connected components), diameter $5$,
radius $3$, center has $84$ vertices, minimal degree $2$, maximal
degree $93$, degree sequence $[ 0, 0, 5, 3, 4, 4, 13, 15, 14, 17,
21, 12, 16, 20, 15$, $14, 11,$ $9, 7, 6, 6, 7, 9, 3, 11$ , $4, 1,
2$, $3, 3, 1, 2, 1, 3, 3, 0, 1, 0, 1, 1, 0,$ $0, 0, 0, 1, 0, 0,
0$, $0, 0, 1$, $0, 0, 1, 1, 2, 1$, $0, 0$, $0, 0, 0, 0$, $0, 0, 0,
0, 0$ , $0, 0, 0$ , $0, 0, 0, 1, 1, 0, 0, 0, 0, 0, 0$, $0, 0, 0$,
$0, 0, 0, 0, 0, 0, 0, 1$, $1 ]$.

Forward, backward, disjunctive and conjunctive folding terminal
graph description - $14,$ $17$, $13$, $21$ vertices, respectively.

Retract folding terminal graph description - strongly connected
graph with $254$ vertices.

%\begin{tabular}{|c|c|c|c|c|}
%  \hline
%  % after \\: \hline or \cline{col1-col2} \cline{col3-col4} ...
%  \#vertices & \#edges & connect.& \#weak comp. & \#big str.conn.comp.   \\
%  \hline
%  54 & 151 & 1 & 1 & 13 \\
%  \hline
%\end{tabular}
%
%Forward fold
%
%\begin{tabular}{|c|c|c|c|c|}
%  \hline
%  % after \\: \hline or \cline{col1-col2} \cline{col3-col4} ...
%  \#vertices & \#edges & connect.& \#weak comp. & \#big str.conn.comp.   \\
%  \hline
%  17 & 269 & 16 & 1 & 1 \\
%  \hline
%\end{tabular}
%
%Backward fold
%
%\begin{tabular}{|c|c|c|c|c|}
%  \hline
%  % after \\: \hline or \cline{col1-col2} \cline{col3-col4} ...
%  \#vertices & \#edges & connect.& \#weak comp. & \#big str.conn.comp.   \\
%  \hline
%  17 & 269 & 16 & 1 & 1 \\
%  \hline
%\end{tabular}
%
%Disjunctive fold
%
%\begin{tabular}{|c|c|c|c|c|}
%  \hline
%  % after \\: \hline or \cline{col1-col2} \cline{col3-col4} ...
%  \#vertices & \#edges & connect.& \#weak comp. & \#big str.conn.comp.   \\
%  \hline
%  17 & 269 & 16 & 1 & 1 \\
%  \hline
%\end{tabular}
%
%Conjunctive fold
%
%\begin{tabular}{|c|c|c|c|c|}
%  \hline
%  % after \\: \hline or \cline{col1-col2} \cline{col3-col4} ...
%  \#vertices & \#edges & connect.& \#weak comp. & \#big str.conn.comp.   \\
%  \hline
%  65 & 1139 & 3 & 1 & 1 \\
%  \hline
%\end{tabular}
%
%Retract fold
%
%\begin{tabular}{|c|c|c|c|c|}
%  \hline
%  % after \\: \hline or \cline{col1-col2} \cline{col3-col4} ...
%  \#vertices & \#edges & connect.& \#weak comp. & \#big str.conn.comp.   \\
%  \hline
%  64 & 1133 & 16 & 1 & 1 \\
%  \hline
%\end{tabular}

\subsubsection{Macaque}

Filename - rhesus\_brain\_1.graphml.

Initial graph description - $242$ vertices and $4090$ edges,
strongly connected, underlying undirected graph has vertex
connectivity $1$ ($3$ cutvertices, $8$ separation pairs), diameter
$4$, radius $3$, center has $147$ vertices, minimal degree $1$,
maximal degree $111$, degree sequence $[ 0, 7, 8, 7, 4, 10, 4, 0,
7$, $6$ , $3$, $3, 5, 8, 2, 4, 6, 9, 12, 3, 7, 5, 4$, $7$, $6$,
$5$, $5, 4, 5, 3, 4, 4, 3$, $4, 4, 1, 6, 2, 2, 7, 2, 7, 1$, $2$,
$2, 0, 1, 1, 5, 1, 1, 1, 0, 3, 1, 0, 1, 0, 0$, $1$, $2$, $1$, $1,
1, 1, 0, 1, 2$, $1, 0, 0, 1, 0, 0, 2, 0, 1$, $0$, $0$, $1$, $0, 0,
0, 0, 0, 0, 0, 0, 0, 0, 0, 0, 0$, $0$, $0$, $0$, $0$, $0$, $0$,
$0$, $0$, $0$, $0$, $0, 0, 0, 0, 0, 0, 0, 0$, $1 ]$.

Forward, backward, disjunctive and conjunctive folding terminal
graph description - $16,$ $17$, $16$, $23$ vertices, respectively,
strongly connected.

Retract folding terminal graph description - strongly connected
graph with $205$ vertices.

\subsubsection{Fly}

Filename - drosophila\_medulla\_1.graphml.

Initial graph description - $1781$ vertices and $9735$ edges,
$996$ strongly connected components - one with $785$ vertices, one
with $2$ vertices, the other components trivial, underlying
undirected graph is disconnected, has $6$ connectivity components
(one big component - $1770$ vertices, connectivity $1$, $265$
cutvertices, diameter $6$, radius $3$, center has $1$ vertex),
minimal degree $1$, maximal degree $927$.

Forward, backward, conjunctive and disjunctive folding terminal
graph description - $334$, $658$, $26$, $1781$ vertices,
respectively.

\subsection{Conclusion}
We introduce study of graph homomorphisms of connectome graphs and
present some initial computation results. For graphs having less
than $2000$ vertices it is possible to compute basic invariants
characterizing graph structure in several minutes on a typical
$2010$s laptop computer. A transition to quotient graphs (as well
as a transition to subgraphs) may be helpful for understanding
operation of nervous systems.

Some of our observations (related to both initial connectome
graphs and terminal graphs):
\begin{enumerate}

\item most connectomes graphs have one big strongly connected
component (SCC) (containing a directed cycle) and an acyclic part,
there are strongly connected examples;

\item vertex connectivity of connected components of underlying
undirected graphs varies, connectivity is higher for graphs
denoted as "brain", connectivity of terminal graphs is bigger or
equal to the initial connectivity;

\item diameter of connected components of underlying undirected
graphs is predominantly small (at most $6$), terminal graphs have
diameter smaller or equal to the initial diameter;

\item folds reduce the size of big SCC, the number of SCC changes
insignificantly (stays constant or increases by $1$);

\end{enumerate}

Further work can be done in the following directions:
\begin{enumerate}

\item interpret the computational results in biological and
modelling terms;

\item introduce constraints or preferences for possible fold types
and folding sequences using additional arguments about graph
structure, edge weights and labels as well as other biological
information, for example, we can first identify pairs of vertices
which has the maximal number of common positive or negative
neighbours;

\item study edge folds;

\item relate the known properties of "rich club" (various
centrality invariants) in terms of terminal graph structures;

\item study structure of strongly connected components of
connectomes and their images under homomorphisms.

%\item study maximal acyclic subgraphs of connectomes and their
%images;

%\item study graph-theoretic invariants of underlying undirected
%graphs of connectomes (e.g. connectivity, diameter, radius and
%other metric invariants)
\end{enumerate}

%%%%%%%%%%%%%%%%%%%%%%%%%%%%%%%%%%%%%%%%%%%%%%%%%%%%%%%%%%%%%

\end{document}